 \documentclass[pmlr,twocolumn,10pt]{jmlr} 

\usepackage{tcolorbox}

\usepackage{booktabs}

\usepackage{siunitx}

\usepackage[switch]{lineno}

\theorembodyfont{\upshape}
\theoremheaderfont{\scshape}
\theorempostheader{:}
\theoremsep{\newline}

\jmlrworkshop{Machine Learning for Health (ML4H) 2025} 

\title{Patient Safety Risks from AI Scribes: \\~Signals from End-User Feedback}

\author{%
\Name{Jessica Dai}
\Email{jessicadai@berkeley.edu}\\
\addr University of California, Berkeley
\AND
\Name{Anwen Huang}
\Email{anwen@berkeley.edu}\\
\addr University of California, Berkeley
\AND
\Name{Catherine Nasrallah} \Email{cathy.nasrallah@ucsf.edu}\\
\addr University of California, San Francisco
\AND
\Name{Rhiannon Croci} \Email{rhiannon.croci@ucsf.edu}\\
\addr University of California, San Francisco
\AND\Name{Hossein Soleimani} \Email{hossein.soleimani@ucsf.edu}\\
\addr University of California,  San Francisco
\AND
\Name{Sarah J. Pollet} \Email{sarah.pollet@ucsf.edu}\\
\addr University of California, San Francisco
\AND
\Name{Julia Adler-Milstein} \Email{julia.adler-milstein@ucsf.edu}\\
\addr University of California, San Francisco
\AND
\Name{Sara G. Murray} \Email{sara.murray@ucsf.edu}\\
\addr University of California, San Francisco
\AND
\Name{Jinoos Yazdany} \Email{jinoos.yazdany@ucsf.edu}\\
\addr University of California, San Francisco
\AND
\Name{Irene Y. Chen} \Email{iychen@berkeley.edu}\\
\addr University of California, San Francisco and University of California, Berkeley
}

\usepackage[x11names]{xcolor}

\usepackage{color-edits} 
\addauthor{jd}{Orchid3}

\begin{document}

\maketitle

\begin{abstract}
AI scribes are transforming clinical documentation at scale.
However, their real-world performance remains understudied, especially regarding their impacts on patient safety. 
To this end, we initiate a mixed-methods study of patient safety issues raised in feedback submitted by AI scribe users (healthcare providers) in a large U.S. hospital system. 
Both quantitative and qualitative analysis suggest that AI scribes may induce various patient safety risks due to errors in transcription, most significantly regarding medication and treatment; however, further study is needed to contextualize the absolute degree of risk.\footnote{This study qualifies as exempt and has undergone limited review by The University of California, San Francisco IRB; IRB \#25-43915, reference \#435961.} 
\end{abstract}

\section{Introduction}
\label{sec:intro}

Medical scribing has long been touted as a canonical application for automation by artificial intelligence (AI) systems. Recent advances in AI development have led to a wide range of ``AI Scribe'' products that claim to realize this vision, and which are now entering widespread use across the healthcare system (see, e.g., \cite{tierney2025ambient} for a survey).\footnote{Some examples of companies that provide AI scribe products include Abridge, Ambience, DeepScribe, Freed, Nabla, ScribePT, Suki, Tali, and so on.}  

Given the nascency of these deployments---and ambiguity about about whether AI scribe products are subject to regulation as medical devices \citep{FDACDSguidance}---it is critical to understand the impact of AI Scribe products when applied to real patient encounters. 
However, while several studies have examined physicians' perspectives (e.g., \cite{tierney2025ambient,shah2025physician, duggan2025clinician}), real-world impacts on patient safety are relatively understudied, with most works utilizing simulated ambulatory encounters (e.g., \cite{anderson5255300evaluating, hose2025development}). 

In this work, we seek to understand risks to patient safety by studying feedback submitted by end-users of an AI scribe product 
deployed in a large U.S. hospital system.
Recent work has proposed end-user feedback as a data source for post-deployment evaluation of AI systems
\citep{dai2025aggregated}; this work explores an application of such a proposal to AI scribes.

Our long-term goal is to design and implement an automated system that can identify safety signals in real time as they arise. 
This initial work is an exploratory analysis that seeks to understand the extent to which end-user feedback identifies patient safety problems, and to identify considerations for future research and development of such an automated system. To these ends, we take a mixed-methods approach.
We begin with standard \textit{quantitative} methods to analyze per-encounter feedback (Section \ref{sec:quant}), which suggest the presence of errors with clinically-significant impact on patient safety.
We thus leverage \textit{qualitative} analysis of complementary survey data to understand these issues in more detail (Section \ref{sec:qual}).

Overall, we find that per-encounter feedback delivered at point-of-care indicates a clear presence of potential patient safety concerns, especially regarding medication and treatment. This is corroborated by concerns raised in survey responses. However, further study is needed to understand the absolute degree of risk, especially in light of 
some positive feedback about safety-relevant attributes in survey responses.

\subsection{System setup and data sources}
\label{sec:setup}

AI scribe products offered by Abridge and a handful of additional vendors were made available to ambulatory providers on an opt-in basis.
In this work, we study two main streams of feedback data. 
In Section \ref{sec:quant}, we analyze ambulatory encounters using Abridge between June 1, 2024 and Oct. 9, 2025.\footnote{This manuscript includes additional data compared to the official Machine Learning for Health findings camera-ready, which included data only until June. The results using additional data, reported here, are generally consistent with those from the prior manuscript.} This feedback is delivered at point-of-care within the scribe app, and takes the form of open-ended text about notes generated for a specific encounter.
In Section \ref{sec:qual}, we analyze free-text comments from a provider survey sent to all users of AI scribes (including non-Abridge vendors).
While the two data sources therefore do not cover identical products and users,  
our goal in this work is not to make comparative claims across vendors but rather to consider clinician feedback as a modality for future study.

\section{Quantitative analysis}
\label{sec:quant}
\begin{table*}[t!]
\centering
\small
\renewcommand{\arraystretch}{1.15}
\setlength{\tabcolsep}{6pt}
\begin{tabular}{p{1.8cm} p{1cm} >{\raggedright\arraybackslash}p{5cm} >{\raggedright\arraybackslash}p{7.4cm}}
\toprule
\textbf{Cluster} & \textbf{Freq.} & \textbf{Manual Summary} & \textbf{LLM Summary} \\
\midrule
\textit{Medication Errors } & 72 (18.5\%)
& Medication name misspellings, incorrect medication plans/dosages
& Lack of clear instructions regarding medication titration, misspelled drug names, incorrect dosages, fabricated medication-related details
 \\
\midrule
\textit{Sleep Medicine-Specific} & 67 (17.2\%)
& Missing sleep disorder symptoms, sleep center phone number
& Missing phone numbers for sleep centers, detailed sleep schedules, and accurate documentation of diagnoses and details related to sleep conditions
 \\
\midrule
\textit{Wording and Patient References} & 61 (15.6\%)
& Issues with wording, how patients are referenced, name misspellings
& Misspelled names, incorrect pronouns, overly formal or inappropriate language, lacking social history \\
\midrule
\textit{Discussion Documentation} & 59 (15.1\%)
& Omitting discussion points from surgical discussions, diagnoses discussions, and physical exam findings discussions
& Inadequate documentation of surgical options, risks, benefits, and patient decisions; 
Omitted diagnoses, physical exam findings, and treatment discussions, fabricated information\\
\midrule
\textit{Misattribution and Misidentification} & 37 (9.5\%)
& Misattribution of statements, confusing who was present at the appointment, incorrect past diagnoses
& Confusion between patient and caregiver histories, inaccurate attribution of symptoms, fabricated details; Omissions of diagnoses, family and social history, and nuanced patient discussions \\
\midrule
\textit{Formatting and Organization} & 34 (8.7\%) 
& Misplacing A/P and PEx in HPI, formatting issues, disorganization
& Disorganized or overly concise HPIs, missing information, inappropriate mixing of sections (e.g., PEx in HPI), missing subjective details like caregiver input and nuanced patient history \\
\midrule
\textit{A/P} & 32 (8.2\%)
& A/P missing information and disorganized, HPI information misplaced in A/P
& Redundancy in the A/P sections, suggestions for separating problems by diagnosis codes and bulleted formats\\
\midrule
\textit{HPI, A/P} & 28 (7.2\%)
& Missing diagnoses, details, prior medication history in HPI \newline
A/P including issues not discussed
& Incorrectly transcribed details and diagnoses, fabricated histories, missing critical discussions and plans, misattributions\\
\bottomrule
\end{tabular}
\caption{\small \textit{Quantitative results: Summaries of BERTopic clusters produced on safety-filtered feedback. LLM summaries have been edited for length; see Table \ref{table:filtered-full} in Appendix \ref{app:quant} for full LLM outputs.}}
\label{table:filtered}
\end{table*}

In this section, we study the per-encounter feedback data as described in Section \ref{sec:setup}. 
Since this work is exploratory, we focus on simple, out-of-the-box approaches rather than methodological development. 
Our main approach is topic modeling, using Sentence-BERT embeddings \citep{reimers2019sentence}
with BERTopic \citep{grootendorst2022bertopic}.

In order to focus on patient safety issues, we take a two-phase approach to analyzing the full set of feedback data. In Step 1, we apply the Sentence-BERT/BERTopic pipeline to the entire feedback dataset. 
The outputs of this initial step included some clusters about non-safety issues (e.g., formatting problems, such as conversions between bullet points and paragraphs, or feature requests). Thus, in Step 2, we remove feedback corresponding to those non-safety clusters, and reapply the clustering pipeline on the remaining ``filtered'' entries. 
Cluster outputs were examined manually for coherence. 

Finally, we are also interested in the extent to which modern LLMs are sufficient for this analysis. To that end, we label the clusters found by BERTopic both manually and with an LLM summarizer. 
As a baseline, we also prompt an LLM to generate overall summaries based on the whole collection of feedback. 

In total, the data consists of 50,123 unique encounters over 470 total physicians. Of these entries, 467 (0.93\% of all encounters) contain text feedback, provided by 55 unique physicians. 
Of these, 77 entries were identified as non-safety-related  in Step 1.

\vspace{4pt}\noindent
\textbf{Results: Patient safety.}
Our initial findings suggest the presence of a variety of safety issues. In Table \ref{table:filtered}, we show the safety-related problems identified by our two-phase clustering approach, with clusters annotated both manually and by GPT-4o.

One prominent safety issue among the feedback submitted at point-of-care is
incorrectly recorded medication names, dosages, and instructions (18.5\% of safety-related feedback); for example, the scribe is reported to provide insufficiently detailed instructions on how to taper medication. 
Other concerns include
incorrect and missing history, diagnosis and treatment information in the History of Present Illness (HPI) and Assessment and Plan (A/P) sections.
For instance, the scribe is reported to both incorrectly record patients' existing conditions as well as to hallucinate conditions the patient does not have; together, A/P and HPI issues comprise 15.4\% of all safety-related feedback. 
A final general concern involves the quality of transcription for extended discussions, especially regarding decisions about major medical procedures (e.g., risks and benefits of surgery).  
The remaining clusters include speaker misattribution (9.5\%); missing details in sleep medicine (17.2\%); formatting and language concerns (24.3\%).

\vspace{4pt}\noindent
\textbf{Results: Methodological insights.}
While simple, our methods in this study highlight three takeaways for future methodological development. 

\begin{figure}
    \centering
\includegraphics[width=1\linewidth]{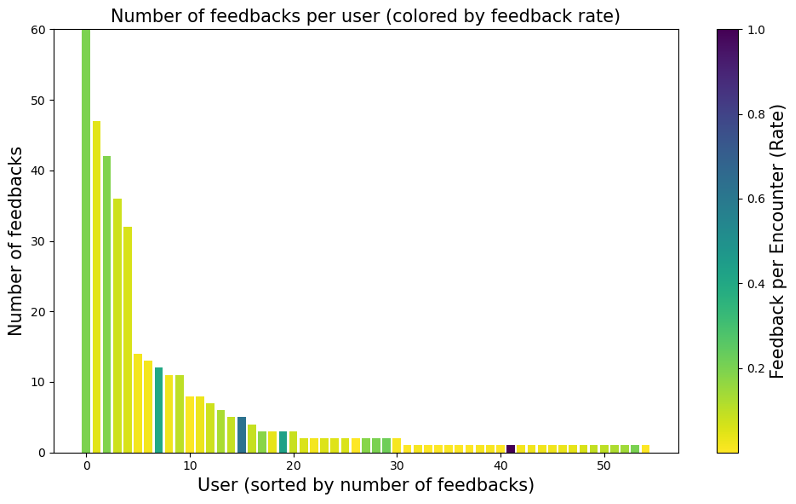}
    \caption{\small \textit{Heterogeneity in feedback volume (bar chart y-axis) and rates (coloring) per physician; each bar corresponds to one unique clinician. y-axis truncated for clarity (top user submitted 151).}}
    \label{fig:hetero}
\end{figure}

First, heterogeneity in per-user behavior is considerable, and shapes the substantive content of feedback overall. For instance, the presence of a sleep-specific cluster appears to reflect one individual sleep medicine specialist who left a high volume of feedback. It is difficult to conclude that the existence of this cluster implies that the scribe struggles especially with sleep medicine contexts, only that such problems were present.
Figure \ref{fig:hetero} illustrates the wide range of feedback volume and rates across the 55 clinicians who submitted text feedback; meanwhile, the overwhelming majority of physicians (415 of 470) submitted no feedback at all. 
See Appendix \ref{app:filtering} for further discussion, and (negative) results for a heuristic attempt at handling outliers. 

Second, the two-step filtering process, while coarse, is an important step given the nature of observed feedback data.
We provide details on this filtering in Appendix \ref{app:filtering}.
In the first iteration of clustering on unfiltered data, 148 out of the 467 feedback entries are marked as noise, despite clearly containing safety-relevant feedback. 
Moreover, while the results of clustering in each step were similar thematically, the Step 2 clusters provided clearer separations in type and location of error. 
The full list of clusters and their respective summaries from the unfiltered dataset can be found in Table \ref{table:unfiltered}.

Finally, LLMs are reasonably capable of providing summaries of short, free-text pieces of feedback, especially after an initial clustering step that ensures some degree of topical coherence. 
We show and discuss these results in Appendix \ref{app:llm-overall}.

\section{Qualitative analysis}
\label{sec:qual}
While encounter-level feedback can provide information about specific examples of scribe errors (or successes), physicians' reflections after repeated usage of scribe products can provide information about their broader impressions of AI scribes overall. 
To this end, clinicians were also given an open-ended survey regarding their usage of AI scribe products. 
A total of 428 physician responses regarding the use of the AI scribe were analyzed, from 279 unique physicians. Of these physicians, 50 were Abridge users, while the remainder used different AI scribe vendors.\footnote{The content of feedback in this survey was very similar across users of different vendors.}

Responses to the open-ended questions were analyzed qualitatively using a multi-stage thematic analysis that incorporated both deductive and inductive approaches \citep{miles1994qualitative}. First, the data were de-identified and reviewed for clarity, relevance, and completeness to inform the development of a preliminary coding framework. Two team members then independently summarized each response and applied a set of inductive codes. Discrepancies were discussed and resolved through consensus meetings with a third coder. Finally, using a grounded approach \citep{braun2006using}, an experienced qualitative researcher reviewed the coded data to generate overarching themes and subthemes.

Thematic analysis revealed several safety-related themes including accuracy, quality and data completeness. Each subtheme included both perceived strengths and limitations of AI scribe integration in clinical practice (Table \ref{table:qual}).

\begin{table*}[t]
\centering
\small
\renewcommand{\arraystretch}{1.1}
\resizebox{\textwidth}{!}{%
\begin{tabular}{p{3cm} p{4.4cm} p{8.0cm}}
\toprule
\textbf{Themes} & \textbf{Positives} & \textbf{Negatives} \\
\midrule
\textit{Accuracy} 
& Error reduction; Accurate output; Focus on medical decision-making; Data synthesis
& Incorrect immunization documentation; Clinical hallucinations regarding diagnoses, exam, and medications; Lack of appropriate clinical reasoning \\
\midrule
\textit{Quality} 
& Patient-friendly language; Minimal editing
& Extensive editing required; Tense inconsistency; Name confusion and misgendering; Speaker recognition errors when multiple people attend visit \\
\midrule
\textit{Data completeness} 
& Comprehensive capture of data
& Missing data for visit components such as history or exam; Incomplete review of systems \\
\bottomrule
\end{tabular}}
\caption{\small \textit{Qualitative results: Themes and sub-themes related to patient safety, with positives and negatives.}}
\label{table:qual}
\end{table*}

\vspace{4pt}\noindent
\textbf{Accuracy.}
Physicians acknowledged the AI scribe’s potential to improve visit documentation and enhance provider efficiency, allowing them to focus more on the patient rather than on documenting patient-reported information. They noted that the AI scribe minimized the likelihood of mistakes and typos that can occur when physicians simultaneously take notes and listen to patients. The AI scribe was recognized in some survey responses for generating accurate clinical documentation, patient information, and AVS, requiring only minor editing. 

However, other feedback expressed notable concerns about the accuracy of the data generated by the AI scribe. Some reported incorrect documentation of immunizations;  others described clinical hallucinations, where the AI scribe fabricated or misrepresented information regarding diagnoses, physical examination findings, symptoms, dates, medication recommendations, and billing details. Furthermore, the lack of appropriate clinical reasoning in some outputs was highlighted by a few physicians as a critical limitation that could negatively impact clinical outcomes.

\vspace{4pt}
\noindent
\textbf{Quality.}
In terms of quality of AI-generated documentation, physicians appreciated the scribe’s ability to generate patient-friendly language, which may enhance the clarity and understanding of clinical notes for patients. In addition, a few physicians noted that the AI-generated notes generally required minimal editing, contributing to more efficient workflows. 

Conversely, several quality-related concerns that could impact patient safety were highlighted by participants. Some physicians described extensive editing as a burden on both their workload and efficiency of work, which was often necessary to correct errors or improve the notes' readability. 
Inconsistencies in verb tense affected overall clarity. 
Physicians also commented on
the scribe’s confusion with patient names and pronouns, as well as voice recognition inaccuracies that led to incorrect documentation. 

\vspace{4pt}
\noindent 
\textbf{Data completeness.}
Some physicians reported that the AI scribe effectively ensured comprehensive capture of clinical data, noting that thorough documentation can support more accurate clinical assessments and continuity of care. 
However, notable gaps in data completeness were also reported by several physicians.
Incomplete documentation 
of the review of systems 
was a frequent challenge, with missing or partially recorded symptoms. Some participants observed missing critical information such as patient names, medical history, and specific test result terminology, which raise concerns about the potential for overlooked clinical details and could compromise the quality of clinical records. 

\section{Discussion}
This work describes initial findings from an analysis of provider feedback about their use of AI scribe products.
To the best of our knowledge, our work is the first to study patient safety risks when AI scribes are used in real-world (non-simulated) encounters; though preliminary, our results suggest the importance of monitoring for, and further study of, patient safety issues in real-world deployments. 

A secondary contribution is in 
identifying per-encounter feedback provided at point-of-care as a potentially-fruitful source of data for conducting such evaluations. However, a substantial amount of this feedback was non-safety related; 
thus, more scaffolding for feedback to focus on safety risks could improve data quality. 
Moreover, as the qualitative results highlight, there is substantial heterogeneity across physicians in their usage and perceptions of the AI scribe. Thus, it is natural to expect that point-of-care feedback also reflects some variation in per-physician behavior (e.g., where some physicians may be more predisposed to submit feedback overall).  

In light of this, we emphasize that these initial results should be thought of as signals that safety issues exist in real-world deployments, rather than definitive claims about absolute degrees of risk, especially given the small sample size. However, a better understanding of true prevalence and severity of risk is necessary. Our results suggest that future methodological work to this end must explicitly handle the nature of one-sided clinician feedback. This might include, e.g., seeking additional data to supplement problems initially raised in clinician feedback, and/or handling behavioral heterogeneity across clinicians. For instance, lack of submitted feedback does not imply lack of safety concerns, as physicians may choose to correct errors without also reporting them. 

While our data is specific to the deployment of AI scribe products within our institution, we believe that the general principles behind the work apply more broadly: ``unknown unknowns'' may persist in any AI deployment, and a key challenge of long-run AI adoption in clinical settings is in developing more mature mechanisms for identifying such issues as they arise. 
We see this work as one step towards this end. 

\newpage
\bibliography{refs}

\newpage
\appendix
\onecolumn

\section{Supplemental materials for quantitative analysis}\label{app:quant}

\subsection{Details on filtering step}
\label{app:filtering}

The initial outputs of our clustering (SentenceBERT+BERTopic) on the full set of data are shown in Table \ref{table:unfiltered}. 
We examined these clusters manually and determined that feedback belonging to clusters 2, 6, 9, and 10 should be removed; these either uninformative, largely positive, or irrelevant to patient safety concerns.

\begin{table}[h!]
\centering
\renewcommand{\arraystretch}{1.2}
\setlength{\tabcolsep}{6pt}
\small
\begin{tabular}{p{1cm}p{0.7cm}p{11cm}}
\hline
\textbf{Cluster} & \textbf{Count} & \textbf{Manual Summary} \\
\hline
-1 & 148 & Mix of issues: misspellings, incorrect discussion transcriptions, making up information, misattribution, misplacing information \\
\hline
0 & 54 & Medication issues: incorrect spelling, dosage, titration schedule \\
\hline
1 & 50 & Sleep medicine related: missing some sleep-specific discussions, resources \\
\hline
2 & 41 & Positive feedback; a couple suggestions on level of detail and transcription issues when an interpreter is present \\
\hline
3 & 40 & Incorrect information: misattribution of symptoms/past illnesses, missing/inaccurate assessment, plan, physical exam \\
\hline
4 & 33 & Mix of positive and negative: negatives focus on misspellings of patient names, confusing patients and physicians, and misgendering patients \\
\hline
5 & 29 & HPI: PEx in HPI, HPI was incorrect/disorganized/not detailed enough \\
\hline
6 & 22 & Positive feedback, specifically about getting the “s” correct in Restless LegS syndrome \\
\hline
7 & 18 & Surgical discussion: inability to capture long surgical discussion about risks and benefits, hallucinations \\
\hline
8 & 18 & Missing diagnoses/incorrect diagnoses in various sections (mostly HPI) \\
\hline
9 & 15 & One-word feedback: “comprehensive”
 \\
\hline
10 & 14 & Sentence fragments (ex. “Hi”, “Also”, “two things:”, “could not use”) \\
\hline
\end{tabular}
\caption{\small \textit{Summaries of BERTopic clusters (pre-filtering)}}
\label{table:unfiltered}
\end{table}

\subsection{Details on handling physician heterogeneity}
\label{app:filtering}

\paragraph{Additional figures illustrating heterogeneity.}

\begin{figure}
    \centering
    \includegraphics[width=0.7\linewidth]{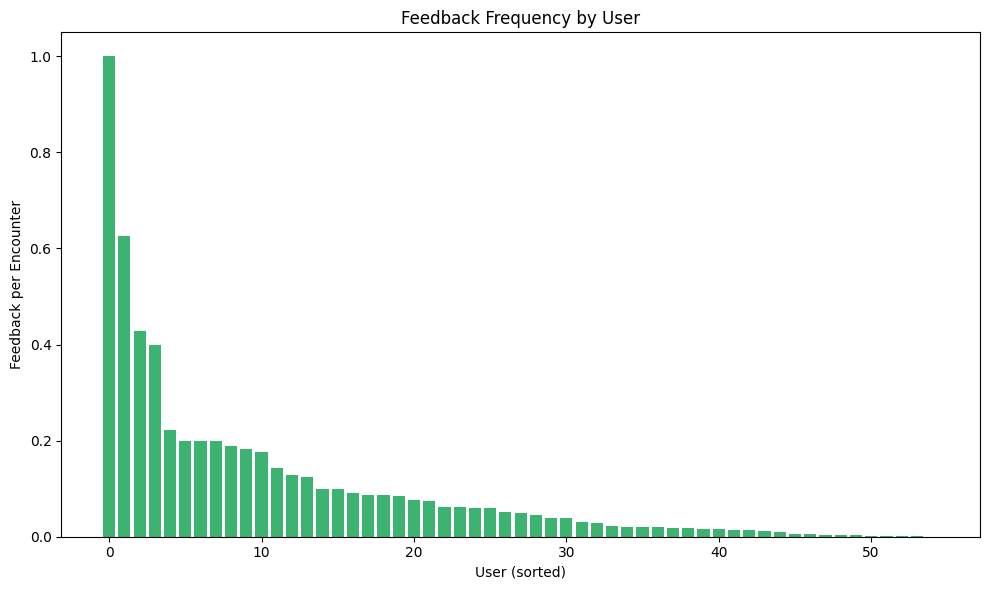}
    \caption{\small \textit{Feedback rates (i.e. fractions of encounters in which feedback was submitted), per user. Alternative visualization of coloring data from Figure \ref{fig:hetero}.}}
    \label{fig:perenc}
\end{figure}

\begin{figure}
    \centering
\includegraphics[width=0.7\linewidth]{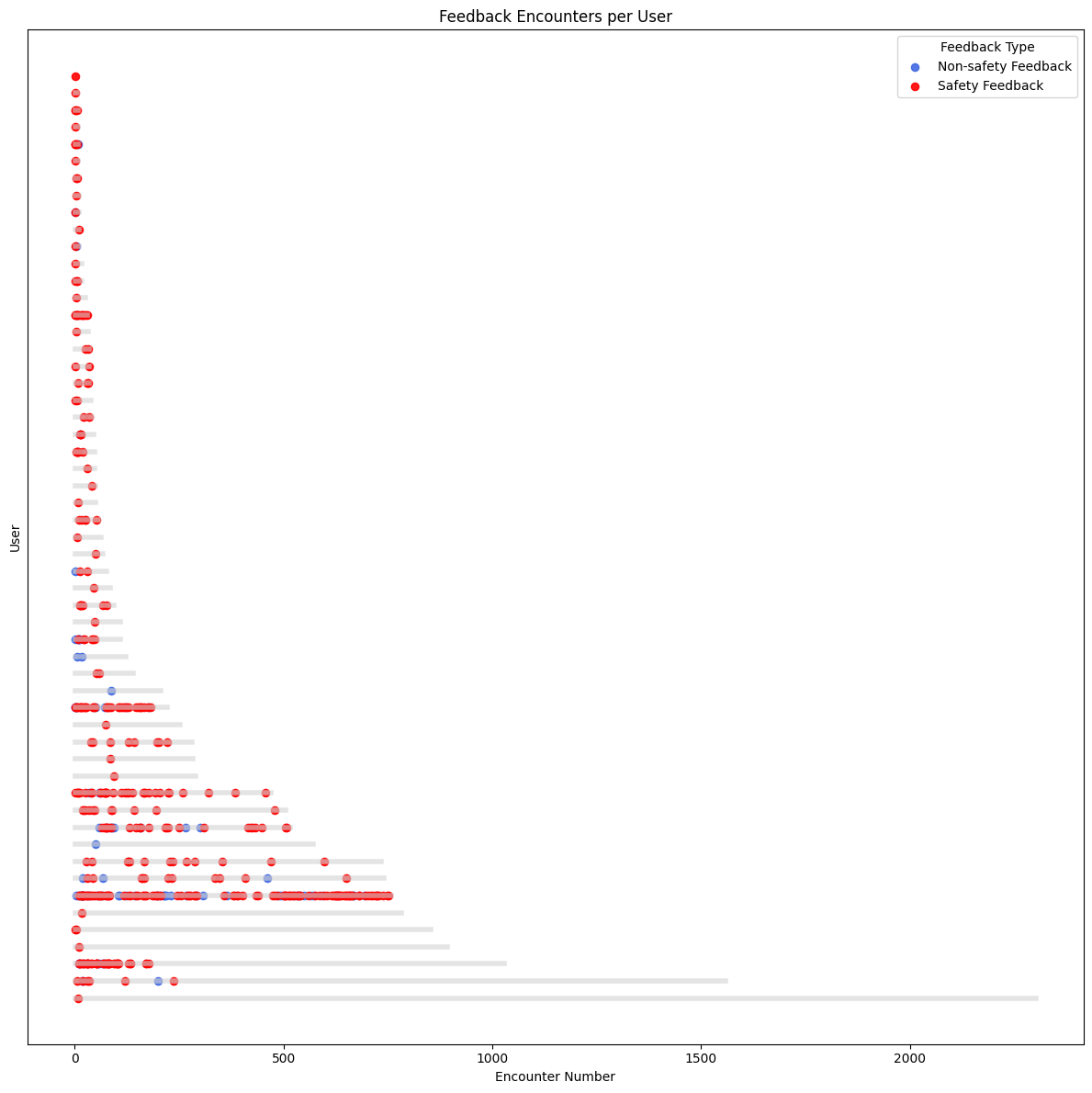}
    \caption{\small \textit{Encounters at which feedback is provided, per user. Blue dots were marked as non-safety related; red were safety-relevant. Gray dots are encounters where no text feedback was provided.}}
    \label{fig:peruser}
\end{figure}

In Figures \ref{fig:perenc} and \ref{fig:peruser}, we provide additional illustrations of per-clinician heterogeneity. Figure \ref{fig:perenc} provides an alternative visualization of the coloring data from Figure \ref{fig:hetero}; on the other hand, \ref{fig:peruser} shows the types of feedback left by users over the course of all encounters with the scribe. In this set of data, the provision of feedback does not appear necessarily to be strongly correlated with any part of the sequence of encounters (e.g., neither only at the begining of encounters nor only at the end of all encounters). 


\paragraph{Condensing feedback volume for outlier physicians.}
Finally, to attempt to handle outlier physicians who submitted a large quantity of feedback, we implemented a feedback capping procedure prior to the second-round clustering. After removing non-safety clusters Step 1, each physician was limited to a maximum of 10 feedback entries. For physicians with more than 10 entries, we computed sentence embeddings of each of their feedback entries, computed the centroid of these embeddings, and then kept the 10 feedback entries with embeddings that were closest in distance (Euclidean) to the centroid. In total, 5 
physicians had their feedback entries capped.
We then performed the second round of clustering on this capped data, the results of which are shown in Table \ref{table:condensed} below. 
However, as the contents of Table \ref{table:condensed} suggest, this procedure did not meaningfully improve the coherence of the final clusters, nor did it dispel the cluster of sleep-specific feedback.

\begin{table}[h!]
\centering
\renewcommand{\arraystretch}{1.2}
\setlength{\tabcolsep}{6pt}
\small
\begin{tabular}{p{1cm}p{0.7cm}p{11cm}}
\hline
\textbf{Cluster} & \textbf{Count} & \textbf{Manual Summary} \\
\hline
-1 & 17 & Issues documenting surgical discussions, inaccurate HPI and A/P \\
\hline
0 & 28 & Mispelling of names, preferences for wording style, inaccurate transcriptions \\
\hline
1 & 22 & HPI issues: formatting, misplaced information, incorrect information \\
\hline
2 & 21 & Incorrect medication names, dosages, and plans \\
\hline
3 & 13 & Omitting details relevant to sleep disorders \\
\hline
\end{tabular}
\caption{\small \textit{Summaries of BERTopic clusters with capped feedback (post-filtering)}}
\label{table:condensed}
\end{table}

\newpage
\subsection{Details on LLM per-cluster summary}
To produce LLM-generated per-cluster summary labels, we concatenated all feedback entries for each cluster into one text chunk, formatted as 
\texttt{Document: [feedback list] | Topic: [cluster label]}. 

We then prompted our internal GPT-4o model with the following prompt, passing in the above list. 

\begin{tcolorbox}[colback=gray!5,colframe=black,boxrule=0.5pt,arc=2pt]
\texttt{You are a precise analyst. Summarize recurring themes per cluster. Here is a list structured as Document: Feedback | Topic label. The topic labels are the different clusters.
}
\vspace{12pt}

\texttt{[feedback and cluster label list]}
\vspace{12pt}

\texttt{Task: Extract 1-2 headline themes for each topic label group (including -1). No more than 2 themes for each.
For each theme, write a short name (2-4 words) and one-sentence description.}
\end{tcolorbox}
The summaries in Table \ref{table:filtered} were edited for space; the full outputs are included in Table \ref{table:filtered-full}.

\begin{table*}[t!]
\centering
\small
\renewcommand{\arraystretch}{1.15}
\setlength{\tabcolsep}{6pt}
\begin{tabular}{p{3cm} p{13.0cm}}
\toprule
\textbf{Cluster} & \textbf{LLM Summary} \\
\midrule
\textit{Medication Errors } 
& Medication Titration Details: Feedback consistently highlights the need for clear and detailed instructions regarding medication titration, including timing, dosage increments, and conditions for changes.\newline
Accurate Transcription for Medications: Issues around misspelled drug names, incorrect dosages, and fabricated medication-related details are frequently flagged, emphasizing the importance of precise transcription.
\\
\midrule
\textit{Sleep Medicine-Specific} 
& Sleep Study and Compliance: Recurring requests focus on including phone numbers for sleep centers, detailed sleep schedules, and accurate documentation of CPAP compliance and benefits.\newline
OSA and Related Diagnoses: Feedback emphasizes ensuring the inclusion of diagnoses related to sleep apnea and related conditions, as well as the detailed discussion of associated topics like BMI and positional aids.
\\
\midrule
\textit{Wording and Patient References} 
& Language and Name Accuracy: Concerns about misspelled names, incorrect pronouns, and overly formal or inappropriate language highlight a need for better contextualization and user-specific preferences.\newline
Diet and Social History: Requests for detailed social and dietary history indicate the importance of these elements, especially in pediatric and patient-specific contexts.
\\
\midrule
\textit{Discussion Documentation} 
& Surgical Discussions and Risk Documentation: Feedback highlights inadequate documentation of surgical options, risks, and benefits, as well as missing patient decisions regarding surgery.\newline
Critical Missing Details: Repeated issues with omitted diagnoses, physical exam findings, treatment discussions, and fabricated information stress the need for completeness and accuracy in notes.
\\
\midrule
\textit{Misattribution and Misidentification}
& Confusion in Attribution: Repeated feedback mentions confusion between patient and caregiver histories, inaccurate attribution of symptoms, and fabricated details, which compromise note integrity.\newline
Missed Key Patient Context: Critical omissions, including specific diagnoses, family and social history, and nuanced patient discussions, limit the utility of the notes.
\\
\midrule
\textit{Formatting and Organization} 
& HPI Organization and Detail: Concerns about disorganized or overly concise HPIs, missing information, and inappropriate mixing of sections (e.g., PEx in HPI) emphasize the need for improved structure and detail.\newline
Contextual Subjectivity: Issues arise around missing subjective details, like caregiver input and nuanced patient history, crucial for accurate medical decision-making.
\\
\midrule
\textit{A/P} 
& Repetition and Conciseness: Feedback often flags redundancy and excessive wordiness in the assessment and plan sections, requesting concise, actionable summaries instead.\newline
Improved Problem Organization: Suggestions for separating problems by diagnosis codes and bulleted formats reflect a desire for clearer organization and alignment with physician inputs.
\\
\midrule
\textit{HPI, A/P} 
& Inaccurate History Documentation: Recurrent concerns about incorrectly transcribed details, fabricated histories, and missing critical discussions underline the need for precise capture of patient interviews.\newline
Assessment and Plan Errors: Issues with inaccurate diagnoses, misattributions, and missing plans highlight gaps in the A/P section, making it less reliable for clinical use.
\\
\bottomrule
\end{tabular}
\caption{\small \textit{Summaries of BERTopic clusters produced on safety-filtered feedback (as shown in Table \ref{table:filtered}), with full LLM outputs.}}
\label{table:filtered-full}
\end{table*}

\subsection{Details on LLM-overall summary}
\label{app:llm-overall}
To produce the overall (non-clustered) LLM summary, 
we concatenated all feedback entries into one text chunk. We then prompted GPT-4o model with the following: 

\begin{tcolorbox}[colback=gray!5,colframe=black,boxrule=0.5pt,arc=2pt]
\texttt{You are a precise analyst. Summarize recurring themes. Here are many short feedback notes: 
}
\vspace{12pt}

\texttt{[list of all feedback]}
\vspace{12pt}

\texttt{Task: Extract 5-10 headline themes across the notes. No more than 10 themes.
For each theme, write a short name (2-4 words) and one-sentence description.}
\end{tcolorbox}

The output of using this prompt on the unfiltered per-encounter data is shown in Table \ref{table:llm-overall-unfiltered}, and on the filtered per-encounter data in Table \ref{table:llm-overall-filtered}. While the latter is remarkably consistent with Table \ref{table:filtered}, it appears that the additional coherence induced by filtering on clusters was necessary, as Table \ref{table:llm-overall-unfiltered} is less well-structured.

\begin{table*}[p]
\centering
\renewcommand{\arraystretch}{1.2}
\setlength{\tabcolsep}{8pt}
\small
\begin{tabular}{p{3cm}p{12cm}}
\toprule
\textbf{Topic} & \textbf{Feedback Issues} \\
\midrule
Accuracy of Information & Feedback consistently highlights inaccuracies in transcriptions, including medication names, diagnoses, procedures, and patient histories, leading to concerns over reliability and precision in documentation.
 \\
\midrule
HPI vs. A/P Organization & A recurring issue involves misplacement of content, with assessment and plan (A/P) details often appearing in the history of present illness (HPI), or vice versa, creating confusion and requiring extensive edits.
\\
\midrule
Mising or Fabricated Details & Many notes omit important patient details such as symptoms, treatment discussions, or specific plans, while some include fabricated or irrelevant information, raising concerns about completeness and integrity.
 \\
\midrule
Medication Titration Instructions & Feedback indicates a need for more detailed and standardized language in medication dosage instructions, emphasizing clarity, timing, and patient-specific adjustments.
 \\
\midrule
Pronouns and Terminology & Errors in pronoun usage, overly formal language, and misinterpretation of conversational tone (e.g., “the patient” vs. name/preferred pronouns) are noted, impacting the professionalism and patient-centric tone of notes.
 \\
\midrule
Phone Number Integration & Many providers express frustration with the failure to consistently include dictated phone numbers, particularly for referrals and follow-up, despite their importance for patient care.
\\
\midrule
Bulleted Format Preference & Providers frequently request concise and bulleted formats for history, plans, and patient instructions, to improve readability and reduce redundancy, especially for after-visit summaries (AVS).
 \\
\midrule
Interpreter and Language Challenges & Misinterpretations or incomplete translations during visits using interpreters, especially in non-English languages, are cited as significant issues affecting note accuracy.
 \\
\midrule
Surgical Discussions and Risks & Surgical options, risk-benefit analyses, and alternative treatments are often poorly summarized or omitted entirely, despite their critical importance in patient counseling. \\
\midrule
Template and Diagnosis Integration & Providers prefer better integration with Epic templates and selected diagnoses, with notes accurately reflecting visit diagnoses and ICD-10 codes rather than generating assumptions.
 \\
\bottomrule
\end{tabular}
\caption{\small \textit{Output of GPT-4o when tasked to identify relevant topics from entire set of (unfiltered) feedback.}}
\label{table:llm-overall-unfiltered}
\end{table*}

\begin{table*}[p]
\centering
\renewcommand{\arraystretch}{1.2}
\setlength{\tabcolsep}{8pt}
\small
\begin{tabular}{p{3.5cm}p{12cm}}
\toprule
\textbf{Topic} & \textbf{Feedback Issues} \\
\midrule
HPI and A/P Organization & Recurring issues involve misplacement of content between the HPI, physical exam, and assessment/plan (A/P), with history often included in A/P or vice versa. Clear delineation of sections is needed for improved note accuracy and readability.
 \\
\midrule
Medication Dosing and Titration & Medication adjustments, tapering schedules, and dosing instructions are frequently incomplete, inaccurate, or overly generalized. Specificity and proper phrasing for titration instructions are essential for patient safety and clarity. \\
\midrule
Missing or Fabricated Information & Several notes omit critical details from discussions or fabricate information, such as diagnoses, patient history, or treatment plans. Ensuring fidelity to the conversation is crucial for medical-legal purposes.
 \\
\midrule
Spelling and Terminology Errors & Misspellings of medications, procedures, providers’ names, and medical terminology (e.g., drug names like “Ziopatch” or “Edarbyclor”) are common and require consistent accuracy checks.
 \\
 \midrule
Lack of Detail in HPI & HPI sections are often too concise, missing pertinent details like symptom characteristics, timeline, lifestyle factors, and social history. Comprehensive documentation improves diagnostic and treatment precision.
 \\
\midrule
Poor Surgical Discussions & Surgical options, risks, benefits, and alternatives are inadequately captured, with missing or unclear information about key aspects of surgery-related decision-making.
\\
\midrule
Inconsistent Follow-Up Details & Follow-up instructions, including phone numbers, referrals, and timelines, are inconsistently included or omitted despite being dictated explicitly. Consistent inclusion is critical for continuity of care.
 \\
\midrule
Pronoun and Subject Confusion & Notes frequently confuse pronouns, misattribute statements between patient and caregivers, or fail to reflect the subjective nature of caregiver-reported information. Proper attribution improves clarity. \\
\midrule
AVS and Patient Instructions & After Visit Summary (AVS) sections are overly wordy and repetitive, often mixing plan and instructions. Simplified, actionable language with clear formatting (e.g., bullet points) is preferred. \\
\midrule
Diagnosis and ICD-10 Accuracy & Generated diagnoses sometimes conflict with visit diagnoses or patient chart history, leading to inaccuracies. Aligning diagnoses with physician input and coding is essential for billing and care clarity. \\
\bottomrule
\end{tabular}
\caption{\small \textit{Output of GPT-4o when tasked to identify relevant topics from feedback filtered to include only safety-relevant concerns.}}
\label{table:llm-overall-filtered}
\end{table*}

\end{document}